\documentclass		{appolb} 

\usepackage		{color} 
\usepackage		{amssymb} 
\usepackage		{epsfig} 

\newcommand{\paragraph}[1]	{} 
\newcommand{\indirizzo}[1]	{ \address{#1} } 
\newcommand{\bea} 		{\begin{eqnarray}}
\newcommand{\eea} 		{\end{eqnarray}}
\newcommand{\beann} 		{\begin{eqnarray*}}
\newcommand{\eeann} 		{\end{eqnarray*}}
\newcommand{\labs} 		{\left\vert}
\newcommand{\rabs} 		{\right\vert}

\newcommand{\lrb} 		{\left(}
\newcommand{\rrb} 		{\right)}

\newcommand{\lab} 		{\left\langle}
\newcommand{\rab} 		{\right\rangle}

\def\pacs		{\PACS}

\begin{document}
\title 			{CONDUCTANCE OF A MOLECULAR WIRE
			\\ ATTACHED TO MESOSCOPIC LEADS:
			\\ CONTACT EFFECTS
%\thanks { Presented at 
% XXIV International School of Theoretical Physics:
% ``Transport Phenomena from Quantum to 
% Classical Regimes'' 
% (September 2000; Ustro{\'n}, Poland)
% }
 			}
\author 		{G. Cuniberti, 
			 G. Fagas, 
			 and K. Richter
\indirizzo 		{Max-Planck-Institut f{\"u}r 
			Physik komplexer Systeme, 
 			\\ N{\"o}thnitzer Stra{\ss}e 38, 
			D--01187 Dresden, Germany
 			}
 			}
\maketitle
%\date{October 30, 2000}
%\date{\today}
\begin{abstract}
We study linear electron transport through a molecular wire sandwiched between nanotube leads.  We show that the presence of such electrodes strongly influences the calculated conductance.  We find that depending on the quality and geometry of the contacts between the molecule and the tubular reservoirs, linear transport can be tuned between an effective Newns spectral behavior
and a more structured one. 
The latter strongly depends on the topology of the leads.  We also provide analytical evidence for an anomalous behavior of the conductance as a function of the contact strength.
\end{abstract}
\pacs { 
 73.50.--h,
 73.61.Wp,
 85.65.+h
 }

The success of semiconductor industry is evidently represented by the empirical exponential ``law'' for the transistors density as a function of time (known as Moore's law~\cite{Moore65}).
Its extrapolation would predict an atomic size gate length in less than two
decades.
The length domain in between is manifestly quantum mechanically dominated:
here conventional methods adopted to characterize electronic devices~\cite{Fischietti84} have to be updated to
more sophisticated ones, including a microscopic treatment of
electronics at the molecular scale~\cite{Taur97}.

The interest in the basic mechanisms of conduction across molecular junctions bridging metallic pads, already object of scanning tunneling microscope oriented research, has been intensified by recent experimental achievements.
For example, $I$--$V$ characteristics of a benzene--1,4--dithiol ring~\cite{RZMBT97}, and a poly(G)--poly(C) 30 base pair long double stranded DNA~\cite{PBdeVD00}
have been reported.
In a parallel development the use of networks of carbon nanotubes (CNT) has been the focus of intense experimental and theoretical activity as another promising direction for building blocks of molecular circuits~\cite{RKJTCL00}.
CNT are known to exhibit a wealth of novel properties depending on their nano--metrical diameter, orientation of graphene roll up, which is conventionally characterized by means of the chiral couple $(n,m)$, and whether they consist of a single (single--wall) or many (multi--wall) cylindrical surfaces~\cite{SDD98+McEuen00}.
Recent experiments also confirm the possibility of producing heterojunctions made of nanotubes~\cite{Venema00+PZXWGP00}.
When such junctions involve different materials~\cite{ADX00+dePGWADR99}, the characterization of contacts becomes a fundamental issue. 
This could be the case when CNT are attached to a single molecule or a molecular cluster with a privileged direction along the current flow, namely a molecular wire. 
CNT have also been recently employed in similar configurations for enhancing the resolution of scanning probe tips~\cite{WJWCL98+NKANHYT00}. 
In such systems the traditional picture of electron transfer between continuum state donor and acceptor species has to be reconsidered in view of the fact that the electrodes are mesoscopic and  the molecule can bear an electric current~\cite{Nitzan01}.

In this paper we drop the conventional assumption of a continuum of (free or quasi--free) lead states. This is a reasonable assumption for mimicking the presence of large reservoirs provided by bulky electrodes, but it may be inadequate when 
the lead lateral dimensions are of the order of the bridged molecule. 
This is indeed the case for CNT.

In what follows, we assume a homogeneous tight binding chain as a model
molecular wire to isolate contact effects. 
Electrodes are taken as either
square lattice tubes (SLT), obtained by imposing periodic boundary conditions on the longitudinal cuts parallel to the lattice bonds, 
or CNT.
The latter consist of a rolled stripe of a graphene honeycomb lattice, also parallel to the lattice bonds. This configuration generates armchair 
single--wall CNT, $(\ell, \ell)$.
%The bond--length is fixed to $a$. 
The consideration of SLT delivers additional analytical insight to the numerically studied model of CNT~\cite{FCR01}.

The electronic Hamiltonian $H = H_\mathrm{tubes} + H_{\rm wire} + H_{\rm coupling}$ of the system reads
\begin{eqnarray}
\label{eq:hamiltonian}
 H &=& 
 \sum_{\alpha={\rm L,R,wire}} \; 
 \sum_{n^{\phantom{\prime}}_{\alpha},n_{\alpha}^\prime} 
 \left( \frac{\varepsilon^{\alpha}_{n_{{\alpha}}}} 
 2 \delta_{n^{\phantom{\prime}}_{{\alpha}},n_{{\alpha}}^\prime}
 -
 \gamma^\alpha_{\left\langle 
 n^{\phantom{\prime}}_{\alpha},n_{\alpha}^\prime \right\rangle}
 \right)
\left| n^{\phantom{\prime}}_{\alpha} \right\rangle 
\left\langle n_{\alpha}^\prime \right| 
 \\ && 
 - \sum_{m_{\rm L} \le M_{\rm c}} 
 %\Gamma_{m_{\rm L}}
 \Gamma
\left| m_{\rm L}\right. \left.\!\! \rangle 
\langle n_{\rm wire}\!=\! 1\right.\left.\!\! \right|
 \nonumber
 - \sum_{m_{\rm R} \le M_{\rm c}} 
 %\Gamma_{m_{\rm R}} 
 \Gamma
\left| m_{\rm R}\right. \left. \! \! \rangle 
\langle n_{\rm wire}\!=\! N\right.\left.\!\! \right|
 + {\rm H.c.} \nonumber
\end{eqnarray}
Here, $\gamma^{\rm L,R}$, $\gamma^{\rm wire}$, and $\Gamma$ are nearest neighbour hopping terms between atoms of the left (L) or right (R) leads, molecular bridge, and the bridge/lead interface, respectively;
$\varepsilon^{\rm wire}$ is the on--site or orbital energy of each of the $n_{\rm wire}=1,\dots, N$ chain--atoms relative to that of the leads, $\varepsilon^{\rm L,R}$.
Note that $n_{\rm L,R}$ is a two--dimensional coordinate spanning the tube lattice.
Summations over $m_{\rm L}$ and $m_{\rm R}$ run over interfacial end--atoms of the leads.
In general, there are $M$ such atomic positions depending on the perimeter of the tubes, whereas, the number of hybridization contacts ranges from 
a single contact (SC) $M_{\rm c}=1$, to multiple contacts (MC) $M_{\rm c}=M$.
 
In order to derive transport properties, we make use of the Landauer theory~\cite{IL99} which relates the conductance of the system to an independent--electron scattering problem~\cite{FG99}.
The electron wavefunction is assumed to extend coherently across the device and the two--terminal, linear--response conductance $g$ at zero temperature 
is simply proportional to the total transmittance for injected electrons 
at the Fermi energy $E_{\rm F}$.
That is, $g=(2 e^2/h)T(E_{\rm F})$, where 
$T=\sum_{j_{\rm L} j_{\rm R}} \labs S_{j_{\rm L} j_{\rm R}} \rabs^2$ 
is a sum over scattering matrix elements labelled by open channels in the leads $j_{L,R}$.
The factor two accounts for spin degeneracy. The transmission function can be calculated from the knowledge of the molecular energy levels, the nature, and the geometry of the contacts. 
It can be obtained by solving the Lippman--Schwinger equation for $H$
and can be written in the following convenient form~\cite{Nitzan01,OKM00+HRHS00},
\bea
\label{eq:transm-mit-spectral-densities}
T(E) = 
4 \Delta^{\rm L}_1 (E) \Delta^{\rm R}_N (E) \labs G_{1N} \lrb E \rrb \rabs^2 ,
\eea
where $G_{ 1 N}$ is the Green function element connecting the two ends of the  $N$--atom--molecule and $\Delta^{L (R)}$ is the left (right) lead spectral density.
The latter is related to the semi--infinite lead Green function matrix ${G}_{\rm lead}$ and is minus the imaginary part of the self--energy 
\bea
\label{eq:spectr-density}
\Sigma^{\alpha=L,R}_{n_{\rm wire}} = \sum_{n_{\alpha}, n_{\alpha}^\prime} 
 \lab n_{\rm wire} | H | n_{\alpha} \rab
 {G}_{\rm lead} \lrb n_{\alpha} , n_{\alpha}^\prime \rrb
 \lab n_{\alpha}^\prime | H | n_{\rm wire} \rab .
\eea
The calculation of the spectral function $\Sigma^{\rm L,R}_{n_{\rm wire}}$
simplifies due to the form of the interfacial coupling in our model.
Assuming identical electrodes, lead-indices are dropped.
Hence, the self--energy becomes 
\beann
\Sigma = 
\frac{\Gamma^2_{\rm eff}}{M_{\rm c}}
\sum_{m, m^\prime \le M_{\rm c}} 
% \Gamma_m \Gamma_{m^\prime}
 {G}_{\rm lead} \lrb m,m^\prime \rrb ,
\eeann
where only surface terms enter in the sum over the states in the leads and
the effective coupling is defined as $\Gamma_{\rm eff}=\Gamma \sqrt{M_{\rm c}}$.

\begin{figure}[t]
\centerline{\epsfig{file=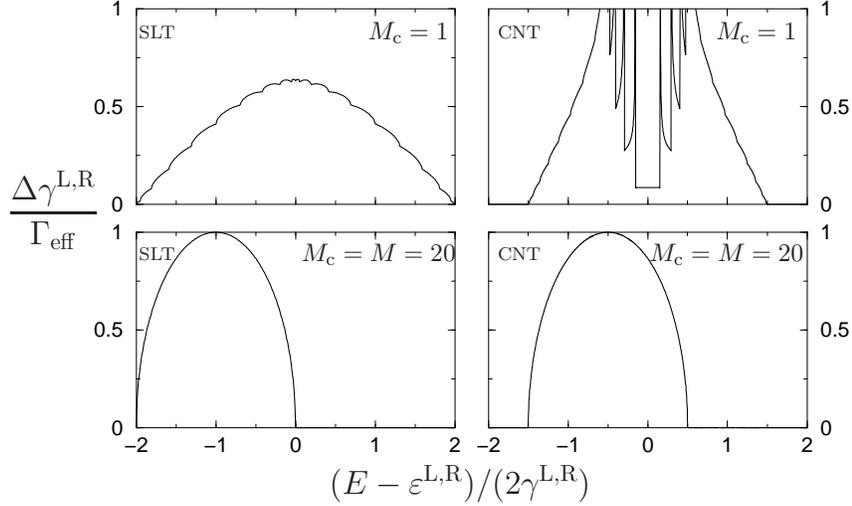, width=0.90\linewidth}}
\caption{\label{fig:fig3} Spectral density plots for square lattice tubes with circumference of $M=20$ atoms (left panels) and $(10,10)$ armchair carbon nanotubes (right). The two different scenarios of a single contact, $M_{\rm c}=1$ (top), and of multiple contacts, $M_{\rm c}=M=20$ (bottom), are shown.}
\end{figure}

In Fig.~\ref{fig:fig3} 
the spectral function $\Delta = - {\rm Im} \Sigma$ is
plotted for both SLT and CNT leads in the SC and MC multiple contact configurations. 
As a function of the number of contacts $M_{\rm c}$ the system interpolates
two different scenarios. In the MC case the spectral density is effectively
the spectral density of one--dimensional leads, obtained by Newns in
his theory of chemisorption~\cite{Newns69}. 
Only the channel without modulation in the transverse profile of its
wavefunction contributes to transport~\cite{CFR00b}. The two--dimensional character of the leads enters as an energy shift of $2 \gamma^{L,R}$ for SLT and of $\gamma^{L,R}$ for CNT, yielding an asymmetric density profile with respect to the atom on--site energy $\varepsilon^{L,R}$.
In contrast, the SC spectrum is symmetric and richer due to the
contribution of all available channels. 
Additional features characterize CNT leads.
The minima in $\Delta$ (upper right panel of Fig.~\ref{fig:fig3}) are responsible for
antiresonances observed in the conductance spectrum~\cite{FCR01}.

In eq.~(\ref{eq:transm-mit-spectral-densities}) $G_{ 1 N}$ can be expressed in the following form~\cite{MKR94b}
\beann
\frac{1}{\gamma^{\rm wire} G_{1N}} = \frac{1}{\zeta(N)} - 
2 \frac{\Sigma}{\gamma^{\rm wire}}
\frac{1}{\zeta(N-1)} + 
\lrb \frac{\Sigma}{\gamma^{\rm wire}} \rrb^2
\frac{1}{\zeta(N-2)} ,
\eeann
where $\zeta(N) / \gamma^{\rm wire}$ is the matrix element $G_{1N}$ for the isolated
$N$--atom molecule. It is given in closed form as a function of the normalized
energy ${\cal E} = (E - \varepsilon^{\rm wire} )/ (2 \gamma^{\rm wire})$ and $N$ by
\beann
\zeta_{\cal E}(N) = \frac { 2 \sqrt{{\cal E}^2 -1}}{\lrb {\cal E} + \sqrt{{\cal E}^2 -1} \rrb^{N+1} - \lrb {\cal E} - \sqrt{{\cal E}^2 -1} \rrb^{N+1}} .
\eeann
In the limit of weak contact coupling the behavior of the $G_{ 1 N}$ element
is dominated by $\zeta (N)$ leading to $N$ transmission resonances in
the conductance of unit height. 
Nevertheless, if the effective coupling between the molecule
and the lead is much larger than $\gamma^{\rm wire}$, $\zeta (N-2)$ will become the dominant term.
As a consequence the conductance spectrum is effectively that of an $(N-2)$--atomic wire~\cite{FCR01}.

In conclusion, we have pointed out that the conductance of molecular wires attached to mesoscopic leads strongly depends on the geometry and dimensionality of the contacts.
In general, a detailed account of the lead states must be provided for describing transport. However, in the multiple contact mesoscopic limit conductance becomes independent of the topology of the tubular electrodes and transport is effectively one--dimensional.
In addition, we have explicitly demonstrated the previously observed dependence of the effective wire length on the coupling strength.

{\it Acknowledgments:} We would like to thank Markus Porto for  critical
reading of the manuscript.

%\vspace*{-.8cm}

\end{document}